\documentclass{sig-alternate-10pt}
\usepackage[margin=1in]{geometry}
\usepackage{booktabs} 
\usepackage{blindtext}
\usepackage{multirow}
\usepackage[toc,page]{appendix}
\usepackage{enumitem}
\usepackage{tabularx}
\usepackage{caption}
\usepackage{xcolor}
\usepackage[labelformat=empty]{caption}

\usepackage{graphicx}  
\usepackage{url}      
\usepackage{amsmath} 
\usepackage{hyperref}

\begin{document}
	\title{False News On Social Media: A Data-Driven Survey}
	\numberofauthors{2}
	
\author{
	\alignauthor
	Francesco Pierri\\
		\affaddr{
    Politecnico di Milano}\\
    \affaddr{Dipartimento di Elettronica, Informazione e Bioingegneria}\\
\email{francesco.pierri@polimi.it}
\alignauthor
	Stefano Ceri\\
	\affaddr{Politecnico di Milano}\\
	    \affaddr{Dipartimento di Elettronica, Informazione e Bioingegneria}\\
\email{stefano.ceri@polimi.it}
}
\maketitle

\begin{abstract}
In the past few years, the research community has dedicated growing interest to the issue of false news circulating on social networks. The widespread attention on detecting and characterizing deceptive information has been motivated by considerable political and social backlashes in the real world. As a matter of fact, social media platforms exhibit peculiar characteristics, with respect to traditional news outlets, which have been particularly favorable to the proliferation of false news. They also present unique challenges for all kind of potential interventions on the subject. 

As this issue becomes of global concern, it is also gaining more attention in academia. 
The aim of this survey is to offer a comprehensive study on the recent advances in terms of detection, characterization and mitigation of false news that propagate on social media, as well as the challenges and the open questions that await future research on the field. We use a data-driven approach, focusing on a classification of the features that are used in each study to characterize false information and on the datasets used for instructing classification methods. At the end of the survey, we highlight emerging approaches that look most promising for addressing false news. 
\end{abstract}

\section{Introduction}
This section serves as an introduction to the topic of false news on social media; we provide some terminology, describe the social media platforms where false news are most widespread, overview psychological and social factors that are involved, discuss some of the effects on the real world and some open challenges. Finally, we discuss the focus of our survey in comparison with other existing surveys. 

\subsection{Terminology}
In recent years, the terms \textbf{false news} and \textbf{fake news} 
have been broadly and interchangeably used to indicate information which can take a variety of flavors: disinformation, misinformation, hoaxes, propaganda, satire, rumors, click-bait and junk news.
We provide next a list of the definitions encountered in the literature, which is by no means exhaustive. While there is common agreement that these terms indicate deceptive information, we believe that an agreed and precise 
definition is still missing.

Some researchers define \textbf{false news} as news articles that are potentially or intentionally misleading for the readers, as they are verifiable and deliberately false \cite{allcott2017,Shu:2017}. They can represent fabricated information which mimics traditional news content in form, but not in the intent or the organizational process \cite{Lazer18}. It has been highlighted how the neologism \textbf{fake news} is usually employed with a political connotation with respect to the more traditional \textit{false news} \cite{Lazer18,Vosoughi18}. 

\textbf{Misinformation} is defined as information that is inaccurate or misleading \cite{Lazer18}. It could spread unintentionally \cite{onlineWWW2018} due to honest reporting mistakes or incorrect interpretations \cite{hernon1995,donfallis2009}. In contrast,
\textbf{disinformation} is false information that is spread deliberately to deceive people \cite{Lazer18} or promote biased agenda \cite{volkova2017}. 


Similarly to disinformation, \textbf{hoaxes} are intentionally conceived to deceive readers; qualitatively, they are described as {\it humorous and mischievous} (as defined in The Oxford English Dictionary) \cite{kumar2016}.

\textbf{Satirical} news are written with the primary purpose of entertaining or criticizing the readers, but similarly to hoaxes they can be harmful when shared out of context \cite{conroy2015, rubin2016}. They are characterized by humor, irony and absurdity and they can mimic genuine news \cite{rubin2015}. 

\textbf{Propaganda} is defined as information that tries to influence the emotions, the opinions and the actions of target audiences by means of deceptive, selectively omitted and one-sided messages. The purpose can be political, ideological or religious \cite{volkova2017, Volkova2018}.

\textbf{Click-bait} is defined as low quality journalism which is intended to attract traffic and monetize via advertising revenue \cite{volkova2017}.

The term \textbf{junk news} is more generic and it aggregates several types of information, from propaganda to hyper-partisan or conspiratorial news and information. It usually refers to the overall content that pertains to a publisher rather than a single article \cite{wolleyphillips2018}.

Finally, we came across several different definitions for \textbf{rumor}. Briefly, a rumor can be defined as a claim which did not originate from news events and that has not been verified while it spreads from one person to another \cite{sunstein2007,allcott2017,Shu:2017}. As there exists a huge literature on the subject, we refer the interested reader to \cite{Zubiaga2018} for an extensive review.

\subsection{Social media platforms as news outlets}

The appearance of false news on news outlets is by no means a new phenomenon: in 1835 a series of articles published on the New York Sun, known as the {\it Great Moon Hoax}, described the discovery of life on the moon \cite{allcott2017}. Nowadays the world is experiencing much more elaborated hoaxes; social media platforms have favored the proliferation of false news with much broader impact.

Most of nowadays \textbf{news consumption} has shifted towards online social media, where it is more comfortable to ingest, share and further discuss news with friends or other readers \cite{gottfried2016, silverman2016, Shu:2017}. As producing content online is easier and faster, \textbf{barriers} for entering online media industry have dropped \cite{allcott2017}. This has conveyed the dissemination of \textbf{low quality} news, which reject traditional journalistic standards and lack of third-party filtering and fact-checking \cite{allcott2017}. These factors, together with a \textbf{decline} of general trust and confidence in traditional mass media, are the primary drivers for the explosive growth of false news on social media \cite{allcott2017, Lazer18}.

Two main motivations have been proposed as to explain the rise of disinformation websites: 1) a \textbf{pecuniary} one, where viral news articles draw significant advertising revenue and 2) a more \textbf{ideological} one, as providers of false news usually aim to influence public opinion on particular topics \cite{allcott2017}.
Besides, the presence of \textbf{malicious agents} such as bots and trolls has been highlighted as another major cause to the spreading of misinformation \cite{botnature, kumar2018}. 

We refer the interested reader to \cite{allcott2017} for an extensive analysis of various factors explaining the spreading of false news in social media platforms.

\subsection{Human factors}

Aside from the technical aspects of social network platforms, the research community has leveraged a set of psychological, cognitive and social aspects which are considered as key contributors to the proliferation of false news on social media. 

Humans have no natural expertise at distinguishing real from false news \cite{Shu:2017, kumar2016}. Two major psychological theories explain this difficulty, respectively called \textbf{naive realism} and the \textbf{confirmation bias}. The former refers to the tendency of users to believe that their view is the only accurate one, whereas those who disagree are biased or uninformed \cite{naiverealism}. The latter, also called \textit{selective exposure}, is the inclination to prefer (and receive) information which confirms existing views \cite{confirmationbias}.
As a consequence, presenting factual information to correct false beliefs is usually unhelpful and may increase misperception \cite{misperception}.

Some studies also mention the importance of \textbf{social identity theory} \cite{socialtheory} and \textbf{normative social influence} \cite{normativetheory}; accordingly, users tend to perform actions which are socially safer, thus consuming and spreading information items that agree with the norms established within the community.


All these factors are related to a certain extent to the well-known \textbf{echo chamber} (or \textit{filter bubble}) effect, which gives rise to the formation of homogeneous clusters where individuals are similar people, that share and discuss similar ideas. 
These groups are usually characterized by extremely polarized opinions as they are insulated from opposite views and contrary perspectives \cite{sunstein2001, sunstein2007, pariser2011}; it has been shown that these close-knit communities are the primary driver of misinformation diffusion \cite{spreading2016}.

Social technologies amplify these phenomena as a result of \textbf{algorithmic bias}, as they promote personalized content based on the preferences of users with the unique goal of maximizing engagement \cite{Lazer18, onlineWWW2018}. 

\subsection{Effects on the real world}

We can explain the explosive growth of attention on false news in light of a series of striking effects that the world has recently experienced.

\textbf{Politics} indeed accounts for most of the attention on false news, as highlighted in \cite{Vosoughi18}. 
The 2016 US presidential elections have officially popularized the term \textit{fake news} to the degree that it has been suggested that Donald Trump may not have been elected president were it not for the effects of false news (and the alleged interference of Russian trolls) \cite{allcott2017}. Likewise, recent studies have shown that false news have also impacted 2016 UK Brexit referendum \cite{howard2016} and the 2017 France presidential elections \cite{ferrara2017}.

Over and above we may recall the \textbf{finance} stock crisis caused by a false tweet concerning president Obama \cite{obama}, the \textbf{shootout} occurred in a restaurant as a consequence of the Pizzagate fake news \cite{Shu:2017} and the diffused mistrust towards \textbf{vaccines} during Ebola and Zika epidemics \cite{ferrara2016, millman2014}.

\subsection{Challenges}
We mention here a few challenges which characterize the fight against false news on social media, as highlighted by recent research on the subject.

Firstly, false news are deliberately created to deceive the readers and to mimic traditional news outlets, resulting in an \textbf{adversarial scenario} where it is very hard to distinguish true news articles from false ones \cite{Shu:2017,botnature}.

Secondly, the \textbf{rate} and the \textbf{volumes} at which false news are produced overturn the possibility to fact-check and verify all items in a rigorous way, i.e. by sending articles to human experts for verification \cite{botnature}. This also raises concern on developing tools for the \textbf{early detection} of false news as to prevent them from spreading in the network \cite{propagation2018}.


Finally, social media platforms impose limitations \cite{misinformation} on the \textbf{collection} of public data and as of today the community has produced very limited training datasets, which typically do not include all the information relative to false news.

\subsection{Survey Focus}

Aside from a few works appeared in 2015 and 2016 \cite{conroy2015,rubin2015,rubin2016}, we build our survey with a focus on the last two years, as most of the research on false news has developed in 2017 and 2018.
Moreover, we concentrate on a few social networks which attracted most of the research focus: \textbf{Twitter}, \textbf{Facebook} and \textbf{Sina Weibo}\footnote{A popular Chinese microblogging website which is a hybrid between Facebook and Twitter.}. This is mainly due to the public availability of data and the existence of proprietary application programming interfaces (API) which ease the burden of collecting data. {\color{black} As a final remark, we considered works covering solely the English language, as this is the prominent approach in the field.}

\begin{table*}
\resizebox{\textwidth}{!}{%
\centering
\label{table}
\begin{tabular}{cccc}
\textbf{}  & \textbf{Machine Learning}  & \textbf{Deep Learning}  & \textbf{Other techniques}  \\ 
\hline
\textbf{Content features}  & \begin{tabular}[c]{@{}c@{}}Wang et al (2017) [76]\\Horne et al. (2017) [24]\\Perez-Rosas et al. (2018) [46]\\Potthast et al. (2018) [44]\\Fairbanks et al. (2018) [12]\end{tabular} & \begin{tabular}[c]{@{}c@{}}Baird et al. (2017) [6] \\Hanselowski et al. (2017) [20]\\Riedel et al. (2017) [50]\\ Wang et al (2017) [76]\\ Popat et al. (2018) [46] \end{tabular} & \begin{tabular}[c]{@{}c@{}}Fairbanks et al. (2018) [12]\\Hosseinimotlagh et al. (2018) [25]\end{tabular} \\ 
\hline
\textbf{Context features}  & Tacchini et al. (2017) [67] & \begin{tabular}[c]{@{}c@{}}Volkova et al. (2017) [73]\\Wang et al. (2018) [77]\\Wu et al. (2018) [79]\\Liu et al. (2018) [32]\end{tabular} & \begin{tabular}[c]{@{}c@{}}Tacchini et al. (2017) [67]\\Wang et al. (2018) [77]\\Yang et al. (2019) [80]\end{tabular} \\ 
\hline
\begin{tabular}[c]{@{}c@{}}\textbf{Content and context}\\\textbf{features}\end{tabular}& \begin{tabular}[c]{@{}c@{}}Shu et al. (2019) [63]\\Volkova et al. (2018) [72]\end{tabular} & \begin{tabular}[c]{@{}c@{}}Ruchansky et al. (2017) [55]\\Volkova et al. (2018) [72]\end{tabular} & Shu et al. (2019) [63] \\
\hline
\end{tabular}
}%
\caption{Table 1. Comparative description of twenty studies for false news detection, in terms of method and considered features.}
\end{table*}

Since our analysis is focused on the aforementioned social media, issues concerning false news on \textbf{collaborative platforms} such as Wikipedia and Yelp (namely fake reviews, spam detection, etc.) are out of the scope of this survey; we thus refer the reader to \cite{kumar2018} for an overview of related research. We suggest \cite{rumors} for a comprehensive review of the research that focuses, instead, on \textbf{rumors detection and resolution}, as we observed that many aspects are shared with our subject. {\color{black}\textbf{Automated fact-checking} is another related topic; it deals with verification rather than search of false news on social media, and we refer the interested reader to \cite{vlachos2}.}
Finally, we suggest \cite{ferrara2016} to the readers who may be interested in the research on \textbf{social bots}.

\section{Problem Formulation and\\ Methodology}


Our presentation of research about false news on social media is divided into three parts. We first describe a huge body of works whose objective is to detect false news, then we describe works that explain the models of diffusion of false news and finally works that attempt to mitigate their effects. 

We start our survey by considering a variegated landscape of research contributions which focus on the \textbf{detection} of false news. {\color{black}Their taxonomy, presented in Table 1, is based on two aspects: employed technique and
considered features.
}

The problem has been traditionally formulated as a supervised binary classification problem, starting with datasets consisting of labeled news articles, related tweets and Facebook posts which allow to capture different features, from content based ones (text, image, video) to those pertaining to the social context (diffusion networks, users' profile, metadata) and, in some cases, to external knowledge bases (Wikipedia, Google News). Labels carrying
the classification into true and false news are typically obtained via fact-checking organizations or by manual verification of researchers themselves.  Appendix A comparatively describes the datasets used as ground truth for false news classification.

For what concerns the classification method, a wide range of techniques are used, from traditional machine learning (Logistic Regression, Support Vector Machines, Random Forest) to deep learning (Convolutional and Recurrent Neural Networks) and to other models (Matrix Factorization, Bayesian Inference). 


{\color{black}
Section 4 describes the literature which focuses on the \textbf{characterization} of misinformation spreading on social media. This is achieved by reconstructing the diffusion networks pertaining to false news, as resulting from multiple users' interactions on the platforms. 

Finally, Section 5 presents a few works which tackle the problem of \textbf{mitigation} against false news on social media, following recent announcements from major platforms to favor crowd-sourcing initiatives against malicious information \cite{crowd1}.}

\section{False News Detection}

We approach these methods by starting from those contributions which focus only on content-based features; we next describe contributions which consider only the social context and finally those that consider both aspects.

\subsection{Content-based}
{\color{black}In this section we consider research contributions which are content-based, meaning that they analyze solely the textual content of news articles, e.g. body, title, source.}

{\color{black}Stance detection as a helpful first step towards fake news detection was introduced during the 2017 Fake News Challenge Stage 1\footnote{\url{http://www.fakenewschallenge.org}} (FNC-1) organized by \textit{D. Pomerleau et al. (2017)} \cite{fnc} (cf. \ref{fnc}). The goal was to classify the stance of an entire news article relative to its headline, i.e. document-level stance detection. Neural networks are employed by three top-performing systems, respectively Talos (\textit{Baird et al. (2017)} \cite{talos}), Athene (\textit{Hanselowski et al. (2017)} \cite{athene}) and UCL Machine Reading (\textit{Baird et al. (2017)} \cite{UCL}). These models rely on a combination of lexical features, including Bag-of-Words, topic modeling and word similarity features. An extensive analysis of these approaches, with experiments on their ability to generalize on unseen data, is provided by \textit{Hanselowski et al. (2018)} \cite{fnc-retrospective}.

\textit{Wang et al. (2017)} \cite{wang2017} consider a multi-label classification task on the Liar dataset (cf. \ref{liar}), one of the first datasets introduced in the literature. This includes several textual and metadata features, such as the speaker affiliation or the source newspaper, and labels are based on the six degrees of truth provided by the PolitiFact\footnote{\url{https://www.politifact.com/}} fact-checking organization. They solve the classification problem by considering several machine learning and
deep learning methods, from logistic regression to convolutional and recurrent neural networks.
}

A deep textual analysis is carried out in \textit{Horne at al. (2017)} \cite{horne2017}, where authors examine the body and title (cf. \ref{buzzfeed}) of different categories of news articles (true, false and satire), extracting complexity, psychological and stylistic features. They highlight the relevance of each {\color{black}aspect} in distinct classification tasks, using a linear Support Vector Machine (SVM), finally inferring that real news are substantially different from false news in title whereas satire and false news are similar in content. They also apply the Elaboration Likelihood Model \cite{elm} to news categories, and suggest that consuming false news requires little energy and cognition, making them more appealing to the readers.

A neural network model is also presented by \textit{Popat et al. (2018)} \cite{popat2018}, who build a framework to classify true and false claims, {\color{black}and also provide self-evidence for the credibility assessment}. They evaluate their model against some state-of-the-art techniques on different collections of news articles (cf. \ref{declare} and \ref{semeval}) and they show examples of explainable results enabled by the \textit{attention mechanism} embedded in the model, which highlights the words in the text that are more relevant for the classification outcome.

{\color{black}
\textit{Perez-Rosas et al. (2018)} \cite{veronica} produce a dataset of false and true news articles (cf. \ref{amtfakenews}) and consider different sets of linguistic features (extracted from the body of news articles) namely ngrams, LIWC \cite{LIWC}, punctuation, syntax and readability. On top of these features they train a linear SVM classifier}, showing different performances depending on the considered feature. They suggest that computational linguistics can effectively aide in the process of automatic detection of false news.

The goal of \textit{Potthast et al. (2018)} \cite{inquiry2018} is to assess the style similarity of several categories of news, notably hyper-partisan, mainstream, satire and false. The proposed methodology employs an algorithm called {\it unmasking} \cite{unmasking}, which is a meta learning approach originally intended for authorship verification. They carry out several experiments comparing topic and style-based features with a Random Forest classifier and they conclude that, while hyper-partisan, satire and mainstream news are well distinguished, a style-based analysis alone is not effective for detecting false news.

{\color{black}\textit{Fairbanks et al. (2018)} \cite{fairbanks2018} also aim to classify false and true news, using a collection of articles gathered from GDELT\footnote{\url{https://www.gdeltproject.org/}}); labels are manually crawled from a fact-checking website\footnote{\url{https://mediabiasfactcheck.com/}}.} They compare two different models, a content-based one which uses a classifier on traditional textual features and a structural method that applies loopy belief propagation \cite{loopy} on a graph built from the link structure of news articles. {\color{black}The conclusions indicate that by modeling just the text content of articles it is possible to detect bias, but it not possible to identify false news.}

\textit{Hosseini et al. (2018)} \cite{hosseinimotlagh2018} tackle the problem of distinguishing different categories of false news (from satire to junk news), based only on the {\color{black}news content. They employ the Kaggle dataset (cf. \ref{kaggle}), where they consider up to six different labels.} Their approach involves a tensor decomposition of documents which aims to capture latent relationships between articles and terms and the spatial/contextual relations between terms. They further use an ensemble method to leverage multiple decompositions in order to discover classes with higher homogeneity and lower outlier diversity. {\color{black}They outperform other state-of-the-art clustering techniques and are able to correctly identify all categories of fake news.}

\subsection{Context-based}
{\color{black}Here we describe research contributions which are (social) context-based in the sense that they focus on information derived from social interactions between users, e.g. likes, comment and (re)tweets, as to detect fake content.}

\textit{Tacchini et al. (2017)} \cite{tacchini2017} propose a technique to identify false news on the basis of users who \textit{liked} them on Facebook. They collect a set of posts and users from both conspiracy theories and scientific pages and they build a dataset where each feature vector represents the set of users who \textit{liked} a page. They eventually compare logistic regression with a (boolean crowdsourcing) harmonic algorithm for showing that they are able to achieve high accuracy with a little percentage of the entire training data.

\textit{Volkova et al. (2017)} \cite{volkova2017} address the problem of predicting four sub-types of suspicious news: satire, hoaxes, click-bait and propaganda. They start from a (manually constructed) list of trusted and suspicious Twitter news accounts and they collect a set of tweets in the period of Brussels bombing in 2016. They incorporate tweet text, several linguistic cues (bias, subjectivity, moral foundations) and user interactions in a \textit{fused} neural network model which is compared against ad-hoc baselines trained on the same features. They qualitatively analyze the characteristics of different categories of news observing the performances of the model.

\textit{Wang et al. (2018)} \cite{eann2018} propose a multi-modal neural network model which extracts both textual and visual features {\color{black}from Twitter and Weibo conversations} in order to detect false news items. Inspired by adversarial settings \cite{goodfellow2014} they couple it with an \textit{event discriminator}, which they claim is able to remove event-specific features and generalize to unseen scenarios, where the number of events is specified as a parameter. They evaluate the model on two custom datasets, but they compare it with ad-hoc baselines which are not conceived for false news detection.

\textit{Wu et al. (2018)} \cite{tracing2018} instead concentrate on modelling the propagation of messages carrying malicious items in social networks. {\color{black}Therefore they build a custom dataset, reflecting both true and false news, by leveraging the Twitter API and the fact-checking website Snopes\footnote{\url{https://www.snopes.com/}}.}
They first infer embeddings for users from the social graph and in turn use a neural network model to classify news items. To this extent they provide a new model to embed a social network graph in a low-dimensional space and they construct a sequence classifier, {\color{black}using \textit{Long Short-Term Memory} (LSTM) networks \cite{lstm} to analyze propagation pathways of messages. They show that their model performs better than other state-of-the-art embedding techniques.}

Propagation of news items is also taken into account by \textit{Yu et al. (2018)} \cite{propagation2018}, who use a combination of convolutional and \textit{Gated Recurrent Units} (GRU) \cite{gru} to model diffusion pathways as multivariate time series, where each point corresponds to the characteristics of the user retweeting the news, and perform early detection of false news. The method is evaluated on two real-world datasets of {\color{black} sharing cascades (cf. \ref{rumors}) showing better performances than other state-of-the-art-techniques, which were nonetheless originally conceived for rumor resolution.}

The first unsupervised approach {\color{black} to false news detection} is provided in \textit{Yang et al. (2019a)} \cite{unsupervisedshu}, where veracity of news and users' credibility are treated as latent random variables in a Bayesian network model, and the inference problem is solved by means of collapsed Gibbs sampling approach \cite{gibbs}. The method is evaluated on LIAR (cf. \ref{liar}) and BuzzFeedNews (cf. \ref{buzzfeed}) datasets, {\color{black} performing better than other} general truth discovery algorithms, not explicitly designed for false news detection.


\subsection{Content and Context-based}
{\color{black}In this section we describe research contributions which consider both news content and the associated social (context) interactions as to detect malicious information items.}

The contribution of \textit{Ruchansky et al. (2017)} \cite{csi2017} is a neural network model which incorporates the text of (false and true) news articles, the responses they receive in social networks and the source users that promote them. The model is tested on Twitter and Weibo {\color{black}sharing cascades} datasets (cf. \ref{rumors}) and it is evaluated against other techniques conceived for rumor detection. They finally present an analysis of users behaviours in terms of lag and activity showing that the source is a promising feature for the detection.

In \textit{Shu et al. (2017)} \cite{trifn} a tri-relationship among publishers, news items and users is employed in order to detect false news. Overall, user-news interactions and publisher-news relations are embedded using non-negative matrix factorization \cite{nonnegative} and users credibility scores. Several different classifiers are built on top of the resulting features and performances are evaluated on the FakeNewsNet dataset (cf. \ref{fakenewsnet}) against other state-of-the-art information credibility algorithms. Results show that the social context could effectively be exploited to improve false news detection.

\textit{Volkova et al. (2018)} \cite{Volkova2018} focus on inferring different deceptive strategies (misleading, falsification) and different types of deceptive news (propaganda, disinformation, hoaxes). Extending their previous work \cite{volkova2017}, they collect summaries, news pages and social media content (from Twitter) that refer to confirmed cases of disinformation. Besides traditional content-based features (syntax and style) they employ psycho-linguistic signals, e.g. biased language markers, moral foundations and connotations, to train different classifiers (from Random Forests to neural networks) in a multi-classification setting. Final results show that falsification strategies are easier to identify than misleading and that disinformation is harder to predict than propaganda or hoaxes. 


\setcounter{figure}{1}
\begin{figure*}
\centering
\includegraphics[width=\textwidth]{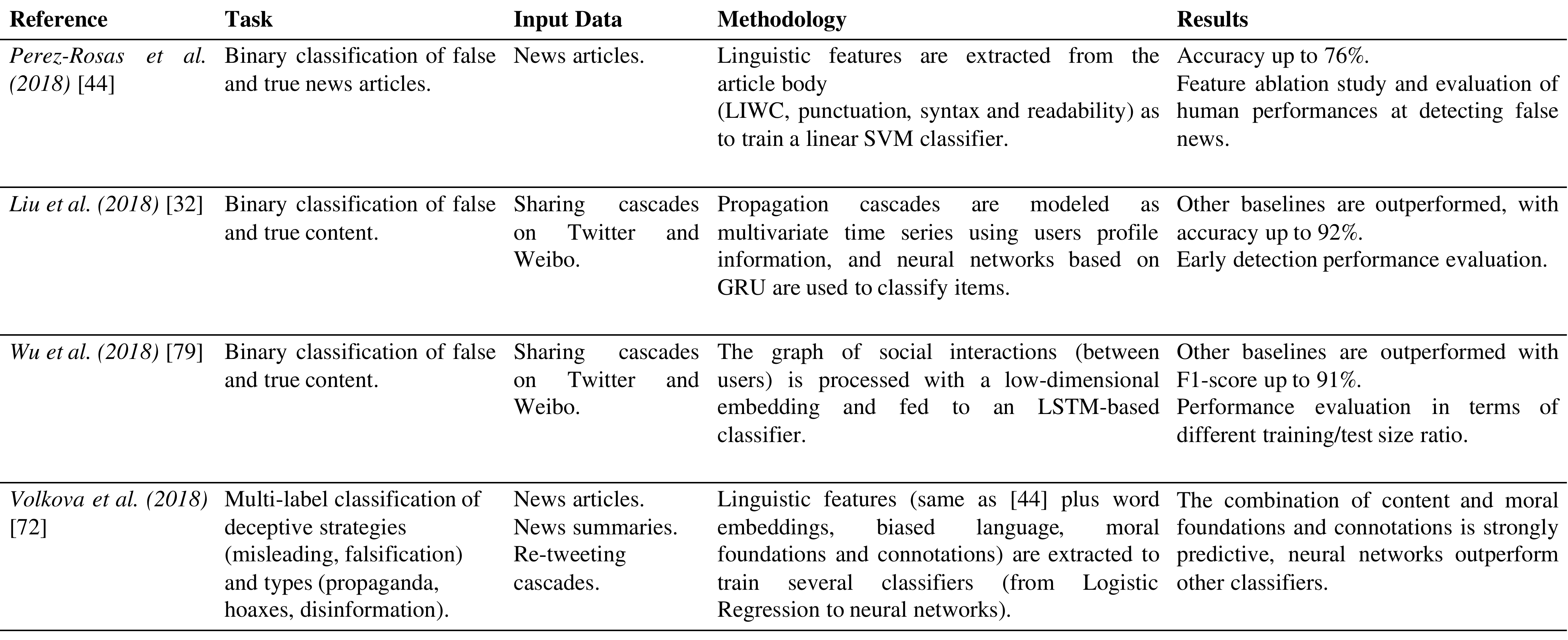}
\vspace{-0.8cm}
\caption*{Table 2: A summary of most promising directions for fake news detection.}
\label{table:table}
\end{figure*}

\setcounter{table}{2}

{\color{black}\subsection{Promising directions}

Despite the vast amount of contributions discussed above, we believe that false news detection requires a deeper and more structured approach. Several works appear as academic exercises, not always compared to each other (and often not comparable).
The main problem of most articles is that they achieve good performance when applied to given input dataset, but they do not generalize to unseen data. From our analysis, it seems that methods purely based upon content analysis work within a limited scope, whereas context analysis addresses generic actions (such as liking, commenting, propagating) that generalize more easily.

Here we highlight most promising approaches among works reviewed so far. They are also summarized in Table 2.



Among the articles from Section 3.1, focused on the content, we cite \textit{Perez-Rosas et al. (2018)} \cite{veronica} for its ability to consider
a huge number of linguistic features, highlighting their different weights on the classification outcome. For such comprehensive approach, this work outstands on approaches based solely on news content; but the approach requires a considerable amount of annotated data (and thus manual efforts) which may hinder the setup of a real-world application.

Among the articles from Section 3.2, focused on the social context, we believe that \textit{Liu et al. (2018)} \cite{propagation2018} and \textit{Wu et al. (2018)} \cite{tracing2018} are most promising; 
they analyzed users' profiles and online news sharing cascades.
Despite the inherent complexity of both techniques (and the limited datasets employed), we argue that a network-based approach focused on social responses might effectively detect deceptive information. They opened the way for new approaches that focus on the models of diffusion of false news on social media, where most of recent research advances stand as described next.

We finally cite \textit{Volkova et al. (2018)} \cite{Volkova2018}, among the articles from Section 3.3 based on both content and context-based features, for considering additional psycho-linguistic signals, e.g. biased language markers, moral foundations and connotations, and inspecting also social responses on Twitter as to infer different deceptive strategies and types of malicious information.
}

\section{Models of False News Diffusion}

A first large-scale study on online misinformation is provided by \textit{Del Vicario et al. (2016)} \cite{spreading2016}, who carry out a quantitative analysis on news consumption relatively to scientific and conspiracy theories news outlets on Facebook. They leverage the Facebook Graph API in order to collect a 5-year span of all the posts (and user interactions) which belong to the aforementioned categories. They analyze cascades (or \textit{sharing trees}) in terms of lifetime, size and edge homogeneity (i.e. an indicator of the polarization of users involved) and they show that 1) the consumption patterns differ in the two categories and that 2) the \textit{echo chambers} (or communities of interest) appear as the preferential drivers for the diffusion of content. On top of these results, they build a data driven percolation model which accounts for homogeneity and polarization and they simulate it in a small-world network reproducing the observed dynamics with high accuracy.

Similarly, a groundbreaking contribution is provided in \textit{Vosoughi et al. (2018)} \cite{Vosoughi18}, where the entire Twitter universe is explored in order to track the diffusion of false and true news. Authors build a collection of links to fact-checking articles (from six different organizations) which correspond to true, false and mixed news stories and they accordingly investigate how these rumors spread on the Twitter network by gathering only tweets that explicitly contain the URLs of the articles. The resulting dataset contains approx. 126000 stories tweeted by 3 million users more than 4.5 million times. A series of measures are carried out including statistical and structural indicators of the retweeting networks along with sentiment analysis, topic distribution and novelty estimation of the different categories of news. The final results show that overall falsehood spread significantly faster, deeper, farther and broader than the truth in all categories of information, with a prominent weight on political news. Moreover, they observe that false news usually convey a higher degree of novelty and that novel information is more likely to be shared by users (although they cannot claim this is the only reason behind the "success" of misinformation). 

A slightly diverse analysis is issued in \textit{Shao et al. (2018a)} \cite{misinformation}, where authors study the structural and dynamic characteristics of the core of the diffusion network on Twitter before and after the 2016 US Presidential Elections. They first illustrate the implementation and deployment of the Hoaxy platform \cite{hoaxy} which is then employed to gather the data required for their analysis. They build different datasets (relative to a few months before and after the elections) which correspond to fact-checking and misinformation articles, i.e. the retweeting network of users that share URLs for related news items, and they perform a k-core decomposition analysis to investigate the role of both narratives in the network. They show that low-credibility articles prevail in the core, whereas fact-checking is almost relegated to the periphery of the network. They also carry out a network robustness analysis in order to analyze the role of most central nodes and guide possible different interventions of social platforms.

Same authors largely extend previous results in \textit{Shao et al. (2018b)} \cite{botnature}, as they carry out a huge analysis on Twitter in a period of ten months in 2016 and 2017. They aim to find evidence of the considerable role of social bots in spreading low-credibility news articles. The Hoaxy \cite{hoaxy} platform is leveraged once again and more than 14 million tweets, including fact-checking and misinformation sources, are collected. The \textit{Botometer} algorithm \cite{botometer} is used to assess the presence of social bots among Twitter users. The results show that bots are active especially in the first phase of the diffusion, i.e. a few seconds after articles are published, and that although the majority of false articles goes unnoticed, a significant fraction tends to become viral. They also corroborate, to a certain extent, results provided by \textit{Vosoughi et al. (2018)} \cite{Vosoughi18}. Moving on, they highlight bot strategies for amplifying the impact of false news and they analyze the structural role of social bots in the network by means of a network dismantling procedure \cite{dismantle}. They finally conclude that 
{\color{black}curbing bots would be an effective strategy to reduce misinformation; using CAPTCHAs \cite{captcha} is a simple tool to distinguish bots from humans, but with undesirable effects to the user experience of a platform.}

Differently from previous works, a study of the agenda-setting \cite{agendasetting} power of false news is instead accomplished in \textit{Vargo et al. (2018)} \cite{agenda2018}, where authors focus on the online mediascape from 2014 to 2016. They leverage a few different agenda-setting models with a computational approach (collecting data from GDELT) in order to examine, among other targets, the influence of false news on real news reports, i.e. whether and to which extent false news have shifted journalistic attention in mainstream, partisan and fact-checking organizations. To this extent they gather news articles corresponding to partisan and mainstream news outlets as well as fact-checking organizations and false news websites; they refer to diverse references in the literature in order to manually construct the list. A network of different events and themes (as identified in the GDELT database) is built to relate distinct media and to model time series of (eigenvector) centrality scores \cite{eigenvector} in order to carry out Granger causality tests and highlight potential correlations. Besides other results, they show that partisan media indeed appeared to be susceptible to the agendas of false news (probably because of the elections), {\color{black}but the agenda setting power of false news--the influence on mainstream and partisan outlets--is declining.}

{\color{black} We described previous works in detail, as we strongly believe that they provided substantial research contributions to the phenomenon of false news spreading on social media, also due to the wide reach of large-scale experiments carried out in these studies. Overall, these approaches have shown that false news spread deeper, faster, broader and farther than the truth, with bots and echo chambers playing a primary role in (dis)information diffusion networks. They also cautiously suggest possible interventions which might be put in place by platform government bodies in order to curb this malicious phenomenon; nonetheless this can not be easily encouraged as it may raise ethical concerns about censorship. As aforementioned, we argue that future research should follow these directions and analyze, from a network perspective, how social communities react to online news as to identify malicious content.}

\section{False News Mitigation}

Finally, a few potential interventions have been proposed for reducing the spread of misinformation on social platforms, from curbing most active (and likely to be bots) users \cite{botnature} to leveraging the users' flagging activity in coordination with fact-checking organizations.
The latter approach is proposed as a first practical mitigation technique in \cite{crowd1} and \cite{crowd2}, where the goal is to reduce the spread of misinformation leveraging users' flagging activity on Facebook.

\textit{Kim et al. (2018)} \cite{crowd1} develop CURB, an algorithm to select the most effective stories to send for fact-checking as to efficiently reduce the spreading of non-credible news with theoretical guarantees; they formulate the problem in the context of temporal point processes \cite{temporal} and stochastic differential equations and they use the Rumors datasets (\ref{rumors}) to evaluate it in terms of precision and misinformation reduction (i.e. the fraction of prevented unverified exposures). They show that the algorithm accuracy is very sensitive to the ability of the crowd at spotting misinformation.

\textit{Tschiatschek et al. (2018)} \cite{crowd2} also aim to select a small subset of news to send for verification and prevent misinformation from spreading; however, as they remark, with a few differences from the previous method respectively 1) they learn the accuracy of individual users rather than considering all of them equally reliable and 2) they develop an algorithm which is agnostic to the actual propagation of news in the network. Moreover, they carry out their experiments in a simulated Facebook environment where false and true news are generated by users in a probabilistic manner. They show that they are able at once to learn users' flagging behaviour and consider possible adversarial behaviour of spammer users who want to promote false news.

A different contribution is issued by \textit{Vo et al. (2018)} \cite{url2018}, who are the first to examine active Twitter users who share fact-checking information in order to correct false news in online discussions. They incidentally propose a URL recommendation model to encourage these \textit{guardians} (users) to engage in the spreading of credible information as to reduce the negative effects of misinformation. They use Hoaxy \cite{hoaxy} (cf. \ref{hoaxy}) to collect a large number of tweets referring to fact-checking organizations and they analyze several characteristics of the users involved (activity, profile, topics discussed, etc). Finally, they compare their recommendation model, which takes into account the social structure, against state-of-the-art collaborative filtering algorithms.

{\color{black}Main social networking platforms, from Facebook to Twitter, have recently provided to their users tools to combat disinformation \cite{crowd1}, an approach which seems reasonable enough to tackle the problem of disinformation without raising censorship alerts. Resorting to the {\it wisdom of the crowd}, as discussed above, can be effective at identifying malicious news items and prevent from misinformation spreading on social networks. }

\section{Conclusions}



Despite the vast review of literature presented so far, in agreement with \cite{Lazer18} we believe that there are only a few substantial research contributions, most of which specifically focus on characterizing the diffusion of misinformation on social media. {\color{black} It has been effectively shown that false news spread faster and more broadly than the truth on social media, and that social bots and echo chambers play an important role in the core of diffusion networks.}

Although different psycho-linguistic signals derived from textual features are useful for false news detection, content alone may not be sufficient and other features, inferred from the social dimension, should be taken into account in order to distinguish false news from true news. 

The lack of gold-standard agreed datasets and of research guidelines on the subject has favored the diffusion of ad-hoc data collections; the related detection techniques share several limitations, as they do not always compare with each other and do not explicitly discuss the impact and consequences of their results.

{\color{black} Nonetheless, the great number of contributions delivered in the last few years shows that the research community has promptly reacted to the issue, and that can successfully embody previous results to advance further in the combat against false news.}

Besides the existing challenges highlighted in the introductory section, we believe that: 1) in light of recent contributions on the characterization of disinformation diffusion networks, {\color{black}more insights into false news detection should be gained from a network perspective;} 2) in general, the research community should coordinate efforts originating from different areas (from psychology to journalism to computer science) in a more structured fashion; 3) future contributions should favor the development of real-world applications for providing effective help in the fight against false news.

\section{Acknowledgements}
F.P. and S.C. are supported by the PRIN grant HOPE (FP6, Italian Ministry of Education). S.C. is partially supported by ERC Advanced Grant 693174.

\bibliography{bib.bib}

\begin{thebibliography}{10}

\bibitem{temporal}
O.~Aalen, O.~Borgan, and H.~Gjessing.
\newblock {\em Survival and event history analysis: a process point of view}.
\newblock Springer Science \& Business Media, 2008.

\bibitem{dismantle}
R.~Albert, H.~Jeong, and A.-L. Barab{\'a}si.
\newblock Error and attack tolerance of complex networks.
\newblock {\em Nature}, 406(6794):378, 2000.

\bibitem{allcott2017}
H.~Allcott and M.~Gentzkow.
\newblock Social media and fake news in the 2016 election.
\newblock {\em Journal of Economic Perspectives}, 31(2):211--36, 2017.

\bibitem{normativetheory}
S.~E. Asch and H.~Guetzkow.
\newblock Effects of group pressure upon the modification and distortion of
  judgments.
\newblock {\em Groups, leadership, and men}, pages 222--236, 1951.

\bibitem{socialtheory}
B.~E. Ashforth and F.~Mael.
\newblock Social identity theory and the organization.
\newblock {\em Academy of management review}, 14(1):20--39, 1989.

\bibitem{talos}
S.~Baird, D.~Sibley, and Y.~Pan.
\newblock Talos targets disinformation with fake news challenge victory.
\newblock {\em Fake News Challenge}, 2017.

\bibitem{gru}
J.~Chung, C.~Gulcehre, K.~Cho, and Y.~Bengio.
\newblock Empirical evaluation of gated recurrent neural networks on sequence
  modeling.
\newblock {\em arXiv preprint arXiv:1412.3555}, 2014.

\bibitem{conroy2015}
N.~J. Conroy, V.~L. Rubin, and Y.~Chen.
\newblock Automatic deception detection: Methods for finding fake news.
\newblock In {\em Proceedings of the 78th ASIS\&T Annual Meeting: Information
  Science with Impact: Research in and for the Community}, page~82. American
  Society for Information Science, 2015.

\bibitem{botometer}
C.~A. Davis, O.~Varol, E.~Ferrara, A.~Flammini, and F.~Menczer.
\newblock Botornot: A system to evaluate social bots.
\newblock In {\em Proceedings of the 25th International Conference Companion on
  World Wide Web}, pages 273--274. International World Wide Web Conferences
  Steering Committee, 2016.

\bibitem{spreading2016}
M.~Del~Vicario, A.~Bessi, F.~Zollo, F.~Petroni, A.~Scala, G.~Caldarelli, H.~E.
  Stanley, and W.~Quattrociocchi.
\newblock The spreading of misinformation online.
\newblock {\em Proceedings of the National Academy of Sciences},
  113(3):554--559, 2016.

\bibitem{semeval}
L.~Derczynski, K.~Bontcheva, M.~Liakata, R.~Procter, G.~W.~S. Hoi, and
  A.~Zubiaga.
\newblock Semeval-2017 task 8: Rumoureval: Determining rumour veracity and
  support for rumours.
\newblock In {\em Proceedings of the 11th International Workshop on Semantic
  Evaluation (SemEval-2017)}, pages 69--76, 2017.

\bibitem{fairbanks2018}
J.~Fairbanks et~al.
\newblock Credibility assessment in the news: Do we need to read?
\newblock In {\em Proc. of the MIS2 Workshop held in conjuction with 11th Int.
  Conf. on Web Search and Data Mining. 799–800.}, 2018.

\bibitem{donfallis2009}
D.~Fallis.
\newblock A conceptual analysis of disinformation.
\newblock {\em iConference}, 2009.

\bibitem{onlineWWW2018}
M.~Fernandez and H.~Alani.
\newblock Online misinformation: Challenges and future directions.
\newblock In {\em Companion of the The Web Conference 2018 on The Web
  Conference 2018}, pages 595--602. International World Wide Web Conferences
  Steering Committee, 2018.

\bibitem{ferrara2017}
E.~Ferrara.
\newblock Disinformation and social bot operations in the run up to the 2017
  french presidential election.
\newblock {\em First Monday}, 22(8), 2017.

\bibitem{ferrara2016}
E.~Ferrara, O.~Varol, C.~Davis, F.~Menczer, and A.~Flammini.
\newblock The rise of social bots.
\newblock {\em Communications of the ACM}, 59(7):96--104, 2016.

\bibitem{vlachos}
W.~Ferreira and A.~Vlachos.
\newblock Emergent: a novel data-set for stance classification.
\newblock In {\em Proceedings of the 2016 conference of the North American
  chapter of the association for computational linguistics: Human language
  technologies}, pages 1163--1168, 2016.

\bibitem{goodfellow2014}
I.~Goodfellow, J.~Pouget-Abadie, M.~Mirza, B.~Xu, D.~Warde-Farley, S.~Ozair,
  A.~Courville, and Y.~Bengio.
\newblock Generative adversarial nets.
\newblock In {\em Advances in neural information processing systems}, pages
  2672--2680, 2014.

\bibitem{gottfried2016}
J.~Gottfried and E.~Shearer.
\newblock {\em News Use Across Social Medial Platforms 2016}.
\newblock Pew Research Center, 2016.

\bibitem{athene}
A.~Hanselowski, P.~Avinesh, B.~Schiller, and F.~Caspelherr.
\newblock Description of the system developed by team athene in the fnc-1.
\newblock {\em Fake News Challenge}, 2017.

\bibitem{fnc-retrospective}
A.~Hanselowski, P.~Avinesh, B.~Schiller, F.~Caspelherr, D.~Chaudhuri, C.~M.
  Meyer, and I.~Gurevych.
\newblock A retrospective analysis of the fake news challenge stance-detection
  task.
\newblock In {\em Proceedings of the 27th International Conference on
  Computational Linguistics}, pages 1859--1874, 2018.

\bibitem{hernon1995}
P.~Hernon.
\newblock Disinformation and misinformation through the internet: Findings of
  an exploratory study.
\newblock {\em Government Information Quarterly}, 12(2):133--139, 1995.

\bibitem{lstm}
S.~Hochreiter and J.~Schmidhuber.
\newblock Long short-term memory.
\newblock {\em Neural computation}, 9(8):1735--1780, 1997.

\bibitem{horne2017}
B.~D. Horne and S.~Adali.
\newblock This just in: fake news packs a lot in title, uses simpler,
  repetitive content in text body, more similar to satire than real news.
\newblock {\em arXiv preprint arXiv:1703.09398}, 2017.

\bibitem{hosseinimotlagh2018}
S.~Hosseinimotlagh and E.~E. Papalexakis.
\newblock Unsupervised content-based identification of fake news articles with
  tensor decomposition ensembles.
\newblock In {\em Proc. of the MIS2 Workshop held in conjuction with 11th Int.
  Conf. on Web Search and Data Mining. 799–800.}, 2018.

\bibitem{howard2016}
P.~N. Howard and B.~Kollanyi.
\newblock Bots,\# strongerin, and\# brexit: computational propaganda during the
  uk-eu referendum.
\newblock {\em arXiv preprint arXiv:1606.06356}, 2016.

\bibitem{crowd1}
J.~Kim, B.~Tabibian, A.~Oh, B.~Sch{\"o}lkopf, and M.~Gomez-Rodriguez.
\newblock Leveraging the crowd to detect and reduce the spread of fake news and
  misinformation.
\newblock In {\em Proceedings of the Eleventh ACM International Conference on
  Web Search and Data Mining}, pages 324--332. ACM, 2018.

\bibitem{junknews}
D.~Liotsiou, B.~Kollanyi, and P.~N. Howard.
\newblock The junk news aggregator: Examining junk news posted on facebook,
  starting with the 2018 us midterm elections.
\newblock {\em arXiv preprint arXiv:1901.07920}, 2019.

\bibitem{unmasking}
M.~Koppel, J.~Schler, and E.~Bonchek-Dokow.
\newblock Measuring differentiability: Unmasking pseudonymous authors.
\newblock {\em Journal of Machine Learning Research}, 8(Jun):1261--1276, 2007.

\bibitem{kumar2018}
S.~Kumar and N.~Shah.
\newblock False information on web and social media: A survey.
\newblock {\em arXiv preprint arXiv:1804.08559, \textit{To appear in the book
  titled Social Media Analytics: Advances and Applications, by CRC press,
  2018}}, 2018.

\bibitem{kumar2016}
S.~Kumar, R.~West, and J.~Leskovec.
\newblock Disinformation on the web: Impact, characteristics, and detection of
  wikipedia hoaxes.
\newblock In {\em Proceedings of the 25th international conference on World
  Wide Web}, pages 591--602. International World Wide Web Conferences Steering
  Committee, 2016.

\bibitem{Lazer18}
D.~M.~J. Lazer, M.~A. Baum, Y.~Benkler, A.~J. Berinsky, K.~M. Greenhill,
  F.~Menczer, M.~J. Metzger, B.~Nyhan, G.~Pennycook, D.~Rothschild,
  M.~Schudson, S.~A. Sloman, C.~R. Sunstein, E.~A. Thorson, D.~J. Watts, and
  J.~L. Zittrain.
\newblock The science of fake news.
\newblock {\em Science}, 359(6380):1094--1096, 2018.

\bibitem{propagation2018}
Y.~Liu and Y.-F. Wu.
\newblock Early detection of fake news on social media through propagation path
  classification with recurrent and convolutional networks.
\newblock {\em AAAI Conference on Artificial Intelligence}, 2018.

\bibitem{rumors}
J.~Ma, W.~Gao, P.~Mitra, S.~Kwon, B.~J. Jansen, K.-F. Wong, and M.~Cha.
\newblock Detecting rumors from microblogs with recurrent neural networks.
\newblock In {\em Proceedings of the Twenty-Fifth International Joint
  Conference on Artificial Intelligence}, pages 3818--3824. AAAI Press, 2016.

\bibitem{agendasetting}
M.~McCombs.
\newblock {\em Setting the agenda: Mass media and public opinion}.
\newblock John Wiley \& Sons, 2018.

\bibitem{millman2014}
J.~Millman.
\newblock The inevitable rise of ebola conspiracy theories.
\newblock {\em The Washington Post}, 2014.

\bibitem{newstrust}
S.~Mukherjee and G.~Weikum.
\newblock Leveraging joint interactions for credibility analysis in news
  communities.
\newblock In {\em Proceedings of the 24th ACM International on Conference on
  Information and Knowledge Management}, pages 353--362. ACM, 2015.

\bibitem{loopy}
K.~P. Murphy, Y.~Weiss, and M.~I. Jordan.
\newblock Loopy belief propagation for approximate inference: An empirical
  study.
\newblock In {\em Proceedings of the Fifteenth conference on Uncertainty in
  artificial intelligence}, pages 467--475. Morgan Kaufmann Publishers Inc.,
  1999.

\bibitem{confirmationbias}
R.~S. Nickerson.
\newblock Confirmation bias: A ubiquitous phenomenon in many guises.
\newblock {\em Review of general psychology}, 2(2):175, 1998.

\bibitem{misperception}
B.~Nyhan and J.~Reifler.
\newblock When corrections fail: The persistence of political misperceptions.
\newblock {\em Political Behavior}, 32(2):303--330, 2010.

\bibitem{elm}
D.~J. O'Keefe.
\newblock Elaboration likelihood model.
\newblock {\em The international encyclopedia of communication}, 2008.

\bibitem{pariser2011}
E.~Pariser.
\newblock {\em The filter bubble: What the Internet is hiding from you}.
\newblock Penguin UK, 2011.

\bibitem{nonnegative}
V.~P. Pauca, F.~Shahnaz, M.~W. Berry, and R.~J. Plemmons.
\newblock Text mining using non-negative matrix factorizations.
\newblock In {\em Proceedings of the 2004 SIAM International Conference on Data
  Mining}, pages 452--456. SIAM, 2004.

\bibitem{LIWC}
J.~W. Pennebaker, R.~L. Boyd, K.~Jordan, and K.~Blackburn.
\newblock The development and psychometric properties of {LIWC}2015.
\newblock Technical report, 2015.

\bibitem{veronica}
V.~P{\'e}rez-Rosas, B.~Kleinberg, A.~Lefevre, and R.~Mihalcea.
\newblock Automatic detection of fake news.
\newblock In {\em Proceedings of the 27th International Conference on
  Computational Linguistics}, pages 3391--3401. Association for Computational
  Linguistics, 2018.

\bibitem{fnc}
D.~Pomerleau and D.~Rao.
\newblock Fake news challenge. http://www.fakenewschallenge.org, 2017.

\bibitem{popat2018}
K.~Popat, S.~Mukherjee, A.~Yates, and G.~Weikum.
\newblock Declare: Debunking fake news and false claims using evidence-aware
  deep learning.
\newblock In {\em Proceedings of the 2018 Conference on Empirical Methods in
  Natural Language Processing}, pages 22--32, 2018.

\bibitem{inquiry2018}
M.~Potthast, J.~Kiesel, K.~Reinartz, J.~Bevendorff, and B.~Stein.
\newblock A stylometric inquiry into hyperpartisan and fake news.
\newblock In {\em Proceedings of the 56th Annual Meeting of the Association for
  Computational Linguistics (Volume 1: Long Papers)}, pages 231--240.
  Association for Computational Linguistics, 2018.

\bibitem{obama}
K.~Rapoza.
\newblock Can 'fake news' impact the stock market?
\newblock {\em Forbes}, 2017.

\bibitem{naiverealism}
E.~S. Reed, E.~Turiel, and T.~Brown.
\newblock Naive realism in everyday life: Implications for social conflict and
  misunderstanding.
\newblock In {\em Values and knowledge}, pages 113--146. Psychology Press,
  2013.

\bibitem{UCL}
B.~Riedel, I.~Augenstein, G.~P. Spithourakis, and S.~Riedel.
\newblock A simple but tough-to-beat baseline for the fake news challenge
  stance detection task.
\newblock {\em arXiv preprint arXiv:1707.03264}, 2017.

\bibitem{kaggle}
M.~Risdal.
\newblock Fake news dataset. https://www.kaggle.com/mrisdal/fake-news.
\newblock 2017.

\bibitem{gibbs}
C.~Robert and G.~Casella.
\newblock {\em Monte Carlo statistical methods}.
\newblock Springer Science \& Business Media, 2013.

\bibitem{rubin2016}
V.~Rubin, N.~Conroy, Y.~Chen, and S.~Cornwell.
\newblock Fake news or truth? using satirical cues to detect potentially
  misleading news.
\newblock In {\em Proceedings of the Second Workshop on Computational
  Approaches to Deception Detection}, pages 7--17, 2016.

\bibitem{rubin2015}
V.~L. Rubin, Y.~Chen, and N.~J. Conroy.
\newblock Deception detection for news: three types of fakes.
\newblock In {\em Proceedings of the 78th ASIS\&T Annual Meeting: Information
  Science with Impact: Research in and for the Community}, page~83. American
  Society for Information Science, 2015.

\bibitem{csi2017}
N.~Ruchansky, S.~Seo, and Y.~Liu.
\newblock Csi: A hybrid deep model for fake news detection.
\newblock In {\em Proceedings of the 2017 ACM on Conference on Information and
  Knowledge Management}, pages 797--806. ACM, 2017.

\bibitem{eigenvector}
B.~Ruhnau.
\newblock Eigenvector centrality: a node centrality?
\newblock {\em Social networks}, 22(4):357--365, 2000.

\bibitem{buzzface}
G.~Santia and J.~Williams.
\newblock Buzzface: A news veracity dataset with facebook user commentary and
  egos.
\newblock {\em International AAAI Conference on Web and Social Media}, 2018.

\bibitem{hoaxy}
C.~Shao, G.~L. Ciampaglia, A.~Flammini, and F.~Menczer.
\newblock Hoaxy: A platform for tracking online misinformation.
\newblock In {\em Proceedings of the 25th International Conference Companion on
  World Wide Web}, WWW '16 Companion, pages 745--750, Republic and Canton of
  Geneva, Switzerland, 2016. International World Wide Web Conferences Steering
  Committee.

\bibitem{botnature}
C.~Shao, G.~L. Ciampaglia, O.~Varol, K.-C. Yang, A.~Flammini, and F.~Menczer.
\newblock The spread of low-credibility content by social bots.
\newblock {\em Nature communications}, 9(1):4787, 2018.

\bibitem{misinformation}
C.~Shao, P.-M. Hui, L.~Wang, X.~Jiang, A.~Flammini, F.~Menczer, and G.~L.
  Ciampaglia.
\newblock Anatomy of an online misinformation network.
\newblock {\em PLOS ONE}, 13(4):1--23, 04 2018.

\bibitem{fakenewsnet}
K.~Shu, D.~Mahudeswaran, S.~Wang, D.~Lee, and H.~Liu.
\newblock Fakenewsnet: A data repository with news content, social context and
  dynamic information for studying fake news on social media.
\newblock {\em arXiv preprint arXiv:1809.01286}, 2018.

\bibitem{Shu:2017}
K.~Shu, A.~Sliva, S.~Wang, J.~Tang, and H.~Liu.
\newblock Fake news detection on social media: A data mining perspective.
\newblock {\em SIGKDD Explor. Newsl.}, 19(1):22--36, Sept. 2017.

\bibitem{trifn}
K.~Shu, S.~Wang, and H.~Liu.
\newblock Beyond news contents: The role of social context for fake news
  detection.
\newblock {\em arXiv preprint arXiv:1712.07709 (2017), to appear in Proceedings
  of 12th ACM International Conference on Web Search and Data Mining (WSDM
  2019)}.

\bibitem{silverman2016}
C.~Silverman.
\newblock This analysis shows how fake election news stories outperformed real
  news on facebook. {BuzzFeed}, \url{https://zenodo.org/record/1239675}, 2016.

\bibitem{sunstein2007}
C.~Sunstein.
\newblock {\em On Rumors: How Falsehoods Spread, Why We Believe Them, What Can
  Be Done.}
\newblock New Haven: Yale University Press. Stowe, 2007.

\bibitem{sunstein2001}
C.~R. Sunstein.
\newblock {\em Echo chambers: Bush v. Gore, impeachment, and beyond}.
\newblock Princeton University Press, 2001.

\bibitem{tacchini2017}
E.~Tacchini, G.~Ballarin, M.~L. Della~Vedova, S.~Moret, and L.~de~Alfaro.
\newblock Some like it hoax: Automated fake news detection in social networks.
\newblock {\em arXiv preprint arXiv:1704.07506}, 2017.

\bibitem{vlachos2}
J.~Thorne and A.~Vlachos.
\newblock Automated fact checking: Task formulations, methods and future
  directions.
\newblock In {\em Proceedings of the 27th International Conference on
  Computational Linguistics}, pages 3346--3359, 2018.

\bibitem{crowd2}
S.~Tschiatschek, A.~Singla, M.~Gomez~Rodriguez, A.~Merchant, and A.~Krause.
\newblock Fake news detection in social networks via crowd signals.
\newblock In {\em Companion of the The Web Conference 2018 on The Web
  Conference 2018}, pages 517--524. International World Wide Web Conferences
  Steering Committee, 2018.

\bibitem{agenda2018}
C.~J. Vargo, L.~Guo, and M.~A. Amazeen.
\newblock The agenda-setting power of fake news: A big data analysis of the
  online media landscape from 2014 to 2016.
\newblock {\em New Media \& Society}, 20(5):2028--2049, 2018.

\bibitem{url2018}
N.~Vo and K.~Lee.
\newblock The rise of guardians: Fact-checking url recommendation to combat
  fake news.
\newblock In {\em The 41st International ACM SIGIR Conference on Research \&
  Development in Information Retrieval}, SIGIR '18, pages 275--284, New York,
  NY, USA, 2018. ACM.

\bibitem{Volkova2018}
S.~Volkova and J.~Y. Jang.
\newblock Misleading or falsification: Inferring deceptive strategies and types
  in online news and social media.
\newblock In {\em Companion Proceedings of the The Web Conference 2018}, WWW
  '18, pages 575--583, 2018.

\bibitem{volkova2017}
S.~Volkova, K.~Shaffer, J.~Y. Jang, and N.~Hodas.
\newblock Separating facts from fiction: Linguistic models to classify
  suspicious and trusted news posts on twitter.
\newblock In {\em Proceedings of the 55th Annual Meeting of the Association for
  Computational Linguistics}, volume~2, pages 647--653, 2017.

\bibitem{captcha}
L.~Von~Ahn, M.~Blum, N.~J. Hopper, and J.~Langford.
\newblock Captcha: Using hard ai problems for security.
\newblock In {\em International Conference on the Theory and Applications of
  Cryptographic Techniques}, pages 294--311. Springer, 2003.

\bibitem{Vosoughi18}
S.~Vosoughi, D.~Roy, and S.~Aral.
\newblock The spread of true and false news online.
\newblock {\em Science}, 359(6380):1146--1151, 2018.

\bibitem{wang2017}
W.~Y. Wang.
\newblock " liar, liar pants on fire": A new benchmark dataset for fake news
  detection.
\newblock In {\em Proceedings of the 55th Annual Meeting of the Association for
  Computational Linguistics (Volume 2: Short Papers)}, volume~2, pages
  422--426, 2017.

\bibitem{eann2018}
Y.~Wang, F.~Ma, Z.~Jin, Y.~Yuan, G.~Xun, K.~Jha, L.~Su, and J.~Gao.
\newblock Eann: Event adversarial neural networks for multi-modal fake news
  detection.
\newblock In {\em Proceedings of the 24th ACM SIGKDD International Conference
  on Knowledge Discovery \& Data Mining}, pages 849--857. ACM, 2018.

\bibitem{wolleyphillips2018}
S.~C. Woolley and P.~N. Howard.
\newblock {\em Computational Propaganda: Political Parties, Politicians, and
  Political Manipulation on Social Media}.
\newblock Oxford University Press, 2018.

\bibitem{tracing2018}
L.~Wu and H.~Liu.
\newblock Tracing fake-news footprints: Characterizing social media messages by
  how they propagate.
\newblock In {\em Proceedings of the Eleventh ACM International Conference on
  Web Search and Data Mining}, pages 637--645. ACM, 2018.

\bibitem{unsupervisedshu}
S.~Yang, K.~Shu, S.~Wang, R.~Gu, F.~Wu, and H.~Liu.
\newblock Unsupervised fake news detection on social media: A generative
  approach.
\newblock In {\em Proceedings of 33rd AAAI Conference on Artificial
  Intelligence}, 2019.

\bibitem{Zubiaga2018}
A.~Zubiaga, A.~Aker, K.~Bontcheva, M.~Liakata, and R.~Procter.
\newblock Detection and resolution of rumours in social media: A survey.
\newblock {\em ACM Comput. Surv.}, 51(2):32:1--32:36, Feb. 2018.

\end{thebibliography}
\bibliographystyle{abbrv}

\appendix
\begin{table*}[!ht]
\centering
\resizebox{\textwidth}{!}{%
\begin{tabular}{ccccccc}
\multicolumn{1}{l}{\begin{tabular}[c]{@{}l@{}} \hphantom{Snopes} \\ \hphantom{whatever} \end{tabular}}& 
\multicolumn{1}{l}{\textbf{Content Features} } & \multicolumn{1}{l}{\textbf{Social Context Features}} & \multicolumn{1}{l}{\textbf{Size}} & \multicolumn{1}{l}{\textbf{Labeling}} & \multicolumn{1}{l}{\textbf{Platform}} & \multicolumn{1}{l}{\textbf{Reference}} \\ \hline
\multicolumn{1}{l}{\textbf{BuzzFeedNews}} & \multicolumn{1}{l}{Article title and source} & \multicolumn{1}{l}{Engagement ratings} & \multicolumn{1}{l}{$10^2$} & \multicolumn{1}{l}{\begin{tabular}[c]{@{}l@{}}\hphantom{NewsTrust}\\ BuzzFeed\\ \hphantom{NewsTrust} \end{tabular}} & \multicolumn{1}{l}{Facebook} & \multicolumn{1}{l}{\cite{silverman2016}} \\ \hline
\multicolumn{1}{l}{\textbf{BuzzFeedWebis}} & \multicolumn{1}{l}{Full Article} & \multicolumn{1}{l}{-} & \multicolumn{1}{l}{$10^3$} & \multicolumn{1}{l}{\begin{tabular}[c]{@{}l@{}}\hphantom{NewsTrust}\\ BuzzFeed\\ \hphantom{NewsTrust} \end{tabular}} & \multicolumn{1}{l}{Facebook} & \multicolumn{1}{l}{\cite{inquiry2018}} \\ \hline
\multicolumn{1}{l}{\textbf{DeClare}} & \multicolumn{1}{l}{Fact-checking post} & \multicolumn{1}{l}{-} & \multicolumn{1}{l}{$10^5$} & \multicolumn{1}{l}{\begin{tabular}[c]{@{}l@{}}NewsTrust\\ PolitiFact\\ Snopes\end{tabular}} & \multicolumn{1}{l}{-} & \multicolumn{1}{l}{\cite{popat2018}} \\ \hline
\multicolumn{1}{l}{\begin{tabular}[c]{@{}l@{}} \hphantom{Snopes} \\ \textbf{FakeNewsAMT} \\ \hphantom{Weibo}\end{tabular}} & \multicolumn{1}{l}{Article text only} & \multicolumn{1}{l}{-} & \multicolumn{1}{l}{$10^3$} & \multicolumn{1}{l}{\begin{tabular}[c]{@{}l@{}}Manual\\ GossipCop \end{tabular}} & \multicolumn{1}{l}{-} & \multicolumn{1}{l}{\cite{veronica}} \\ \hline
\multicolumn{1}{l}{\begin{tabular}[c]{@{}l@{}} \hphantom{Snopes} \\ \textbf{FakeNewsChallenge} \\ \hphantom{Weibo}\end{tabular}} & \multicolumn{1}{l}{Full article} & \multicolumn{1}{l}{-} & \multicolumn{1}{l}{$10^3$} & \multicolumn{1}{l}{\begin{tabular}[c]{@{}l@{}}Manual \end{tabular}} & \multicolumn{1}{l}{-} & \multicolumn{1}{l}{\cite{fnc}} \\ \hline
\multicolumn{1}{l}{\begin{tabular}[c]{@{}l@{}} \hphantom{Snopes} \\ \textbf{FakeNewsNet} \\ \hphantom{Weibo}\end{tabular}} & \multicolumn{1}{l}{Full article} & \multicolumn{1}{l}{Users metadata} & \multicolumn{1}{l}{$10^3$} & \multicolumn{1}{l}{\begin{tabular}[c]{@{}l@{}}BuzzFeed\\ PolitiFact \end{tabular}} & \multicolumn{1}{l}{Twitter} & \multicolumn{1}{l}{\cite{fakenewsnet}} \\ \hline
\multicolumn{1}{l}{\textbf{Hoaxy}} & \multicolumn{1}{l}{Full article} & \multicolumn{1}{l}{\begin{tabular}[c]{@{}l@{}}Diffusion network\\ Temporal trends\\ Bot score (for users)\end{tabular}} & \multicolumn{1}{l}{$>10^6$} & \multicolumn{1}{l}{-} & \multicolumn{1}{l}{Twitter} & \multicolumn{1}{l}{\cite{hoaxy}} \\ \hline
\multicolumn{1}{l}{\textbf{Kaggle}} & \multicolumn{1}{l}{Article text and metadata} & \multicolumn{1}{l}{-} & \multicolumn{1}{l}{$ 10^4$} & \multicolumn{1}{l}{\begin{tabular}[c]{@{}l@{}} \hphantom{BuzzFeed}\\ BS Detector \\ \hphantom{whatever}\end{tabular}} & \multicolumn{1}{l}{-} & \multicolumn{1}{l}{\cite{kaggle}} \\ \hline
\multicolumn{1}{l}{\textbf{Liar}} & \multicolumn{1}{l}{Short statement} & \multicolumn{1}{l}{-} & \multicolumn{1}{l}{$10^4$} & \multicolumn{1}{l}{\begin{tabular}[c]{@{}l@{}} \hphantom{BuzzFeed}\\ PolitiFact \\ \hphantom{whatever}\end{tabular}} & \multicolumn{1}{l}{-} & \multicolumn{1}{l}{\cite{wang2017}} \\ \hline
\multicolumn{1}{l}{\textbf{SemEval-2017 Task8}} & \multicolumn{1}{l}{\begin{tabular}[c]{@{}l@{}}Full article\\ Wikipedia articles\end{tabular}} & \multicolumn{1}{l}{Threads (tweets, replies)} & \multicolumn{1}{l}{$10^4$} & \multicolumn{1}{l}{\begin{tabular}[c]{@{}l@{}} \hphantom{BuzzFeed}\\ Manual \\ \hphantom{whatever}\end{tabular}} & \multicolumn{1}{l}{Twitter} & \multicolumn{1}{l}{\cite{semeval}}  \\ \hline
\multicolumn{1}{l}{\begin{tabular}[c]{@{}l@{}} \hphantom{Snopes} \\ \textbf{Rumors} \\ \hphantom{Weibo}\end{tabular}} & \multicolumn{1}{l}{Fact-checking title} & \multicolumn{1}{l}{\begin{tabular}[c]{@{}l@{}}Diffusion network (Twitter)\\ Original message, replies (Weibo)\end{tabular}} & \multicolumn{1}{l}{$10^4$} & \multicolumn{1}{l}{\begin{tabular}[c]{@{}l@{}} Snopes \\ Weibo\end{tabular}} & \multicolumn{1}{l}{\begin{tabular}[c]{@{}l@{}}Twitter\\ Sina Weibo\end{tabular}} & \multicolumn{1}{l}{\cite{rumors}} \\ \hline
 &  &  &  &  &  & 
\end{tabular}%
}
\centering
\caption{Table 3. Comparative description of the datasets referenced in this survey.}
\end{table*}
\section{Datasets}
The research community has produced a rich but heterogeneous ensemble of data collections for fact checking, often conceived for similar objectives and for slightly different tasks. We first introduce the datasets which are referenced in this survey along with a short description, the source and the main references; their features are summarized in Table 2. Next, we present some other interesting datasets.

\subsection{BuzzFeedNews}
\label{buzzfeed}
BuzzFeed\footnote{\url{https://www.buzzfeed.com}} News journalists have produced different collections of verified false and true news, shared by both hyperpartisan and mainstream news media on Facebook in 2016 and 2017; two of them, introduced by \textit{Silverman (2016)} \cite{silverman2016}, consist of title and source of news items and they are used in \cite{unsupervisedshu, horne2017, buzzface}

\subsection{BuzzFeed-Webis}
\label{buzzfeed-webis}
This collection
extends the previous one as it also contains the full content of shared articles with attached multimedia; it is employed in \cite{inquiry2018}.


\subsection{DeClare}
\label{declare}
This dataset contains several articles from Snopes, PolitiFact and NewsTrust \cite{newstrust} corresponding to both true and false claims; it is proposed in \cite{popat2018} and used for false news detection. 

\subsection{FakeNewsAMT}
\label{amtfakenews}
This collection contains some legitimate articles from mainstream news, some false news generated by Amazon Mechanical Turk workers and some false and true claims from GossipCop\footnote{\url{https://www.gossipcop.com}} (a celebrity fact-checking website); it is introduced in \cite{veronica} for false news detection.

{\color{black}\subsection{FakeNewsChallenge}
\label{fnc}
This dataset was proposed for the 2017 Fake News Challenge Stage 1 \cite{fnc}; it contains thousands of headlines and documents which have to be classified in a document-based stance detection task using 4 different labels (Agree, Discuss, Disagree, Unrelated). It was inspired by \cite{vlachos} where stance detection is instead applied at the level of single sentences. It is employed in \cite{athene, talos, UCL}; an additional analysis is provided in \cite{fnc-retrospective}.}

\subsection{FakeNewsNet}
\label{fakenewsnet}
This dataset contains both news content (source, body, multimedia) and social context information (user profile, followers/followee) regarding false and true articles, collected from Snopes and BuzzFeed and shared on Twitter; it was presented in \cite{fakenewsnet} and employed in \cite{trifn}.


\subsection{Hoaxy}
\label{hoaxy}
The Hoaxy platform\footnote{\url{https://hoaxy.iuni.iu.edu}} has been first introduced in \cite{hoaxy} and employed in several studies \cite{url2018,botnature,misinformation} for different goals; it is continuously monitoring the diffusion network (on Twitter, since 2016) of news articles from both disinformation and fact-checking websites and it allows to generate custom data collections.
\subsection{Kaggle}
\label{kaggle}
This dataset was conceived for a Kaggle false news detection competition \cite{kaggle} which contains text and metadata from websites indicated in the BS Detector\footnote{https://github.com/bs-detector/bs-detector}; it is employed in \cite{hosseinimotlagh2018}.

\subsection{Liar}
\label{liar}
This is a collection of short labeled statements from political contexts, collected from PolitiFact, which serve for false news classification; it first appeared in \cite{wang2017} and it is employed in \cite{unsupervisedshu}.

\subsection{SemEval-2017 Task8}
\label{semeval}
This data collection, composed of tweets and replies which form specific \textit{conversations}, was designed for the specific tasks of stance and veracity resolution of social media content on Twitter; it is described in \cite{semeval} and used in \cite{popat2018}.
\subsection{Rumors}
\label{rumors}
This dataset was originally conceived for rumor detection and resolution in Twitter and Sina Weibo; introduced in \cite{rumors}, it contains retweet and discussion cascades corresponding to rumors/non-rumors and it is employed for false news detection and mitigation in \cite{csi2017, crowd1, propagation2018}.

\subsection{Others}
\textbf{BuzzFace} is a novel data collection composed of annotated news stories that appeared on Facebook during September 2016; it extends previous BuzzFeed dataset(s) (cf. \ref{buzzfeed}) with comments and the web-page content associated to each news article; itjunknews is introduced in \cite{buzzface}. 

As a complement to Hoaxy (cf. \ref{hoaxy}), \textbf{JunkNewsAggregator} is a platform that tracks the spread of disinformation on Facebook pages; it is described in \cite{junknews}. 

Other datasets point to relevant organizations in the context of false news: \cite{agenda2018} contains a list of false news outlets as indicated by different fact-checking organizations, whereas the list of signatories\footnote{\url{https://ifcncodeofprinciples.poynter.org/signatories}} of the International Fact Checking Network's code of principles is a collector of the main fact-checking organizations which operate in different countries. 
Finally, \cite{allcott2017} provides a set of the most shared false articles identified on Facebook during 2016 US elections.

\end{document}